# Fabrication-induced even-odd discrepancy of magnetotransport in few-layer $MnBi_2Te_4$

Yaoxin Li[1,*], Yongchao Wang[1,*], Zichen Lian[1,*], Hao Li[2,3], Zhiting Gao[4], Liangcai Xu[1], Huan Wang[5,6], Rui'e Lu[7], Longfei Li[8], Yang Feng[6], Jinjiang Zhu[9], Liangyang Liu[1], Yongqian Wang[5,6], Bohan Fu[5,6], Shuai Yang[5,6], Luyi Yang[1,10], Yihua Wang[9,11], Tianlong Xia[5,6], Chang Liu[12], Shuang Jia[8,13,14], Yang Wu[15], Jinsong Zhang[1,10,16], Yayu Wang[1,10,16], Chang Liu[5,6†]

[1]*State Key Laboratory of Low Dimensional Quantum Physics, Department of Physics, Tsinghua University, Beijing 100084, China*

[2]*School of Materials Science and Engineering, Tsinghua University, Beijing 100084, China*

[3]*Tsinghua-Foxconn Nanotechnology Research Center, Department of Physics, Tsinghua University, Beijing 100084, China*

[4]*Beijing Academy of Quantum Information Sciences, Beijing 100193, China*

[5]*Beijing Key Laboratory of Opto-electronic Functional Materials & Micro-Nano Devices, Department of Physics, Renmin University of China, 100872 Beijing, China*

[6]*Key Laboratory of Quantum State Construction and Manipulation (Ministry of Education), Renmin University of China, Beijing 100872, China*

[7]*School of Mechanical and Electric Engineering, Guangzhou University, Guangzhou 510006, China*

[8]*International Center for Quantum Materials, School of Physics, Peking University, Beijing 100871, China*

[9]*State Key Laboratory of Surface Physics and Department of Physics, Fudan University, Shanghai 200433, China*

[10]*Frontier Science Center for Quantum Information, Beijing 100084, China*

[11]*Shanghai Research Center for Quantum Sciences, Shanghai 201315, China.*

[12]*Shenzhen Institute for Quantum Science and Engineering and Department of Physics, Southern University of Science and Technology, Shenzhen 518055, China*

[13]*Interdisciplinary Institute of Light-Element Quantum Materials and Research Center for Light-Element Advanced Materials, Peking University, Beijing 100871, China*

[14]*CAS Center for Excellence in Topological Quantum Computation, University of Chinese Academy of Sciences, Beijing 100190, China*

[15]*College of Math and Physics, Beijing University of Chemical Technology, Beijing 100029, China*

[16]*Hefei National Laboratory, Hefei 230088, China*

[*] These authors contributed equally to this work.

[†] Emails: liuchang_phy@ruc.edu.cn

**The van der Waals antiferromagnetic topological insulator $MnBi_2Te_4$ represents a promising platform for exploring the layer-dependent magnetism and topological states of matter. Recently observed discrepancies between magnetic and transport properties have aroused controversies concerning the topological nature of $MnBi_2Te_4$ in the ground state. In this article, we demonstrate that fabrication can induce mismatched even-odd layer dependent magnetotransport in few-layer $MnBi_2Te_4$. We perform a comprehensive study of the magnetotransport properties in 6- and 7-septuple-layer $MnBi_2Te_4$, and reveal that both even- and odd-number-layer device can show zero Hall plateau phenomena in zero magnetic field. Importantly, a statistical survey of the optical contrast in more than 200 $MnBi_2Te_4$ flakes reveals that the zero Hall plateau in odd-number-layer devices arises from the reduction of the effective thickness during the fabrication, a factor that was rarely noticed in previous studies of 2D materials. Our finding not only provides an explanation to the controversies regarding the discrepancy of the even-odd layer dependent magnetotransport in $MnBi_2Te_4$, but also highlights the critical issues concerning the fabrication and characterization of 2D material devices.**

## Introduction

The antiferromagnetic (AFM) topological insulator (TI) $MnBi_2Te_4$ provides promising opportunities for exploring various quantized topological phenomena[1-6]. As a layered *A*-type antiferromagnet, $MnBi_2Te_4$ bulk crystal is composed of septuple layers (SLs) stacked along the *c*-axis with intralayer ferromagnetic (FM) order and interlayer AFM order (Fig. 1a). The interplay between magnetic order and band topology gives rise to gapped surface states that exhibit half-quantized surface Hall conductivity $\sigma_{xy} = 0.5\ e^2/h$, where $h$ represents the Plank constant and $e$ denotes the elementary charge[7,8]. Therefore, depending on the magnetizations of the top and bottom surfaces, few-layer $MnBi_2Te_4$ with different SL-number-parity exhibits distinct topological quantum states[9]. In odd-number-SL $MnBi_2Te_4$, the parallel magnetization on the two surfaces gives rise to the quantum anomalous Hall (QAH) effect[3,10] characterized by quantized Hall resistivity ($\rho_{yx}$) and vanished longitudinal resistivity ($\rho_{xx}$) at zero magnetic field ($H$). In contrast, even-number-SL $MnBi_2Te_4$ displays a robust zero Hall plateau $\rho_{yx} = 0$ and large $\rho_{xx}$ in a wide range of both $\mu_0 H$ and gate voltage ($V_g$), as the counter-propagating Hall currents in the two surfaces cancel out[11-13]. Because the zero Hall plateau with Chern number $C = 0$ is closely related to the topological magnetoelectric effect that stems from the axion electrodynamics[14-16], magnetic TI with antiparallel magnetizations of two surfaces is widely believed as an ideal system for realizing the axion insulator state[5,17-19]. Recently, using a circularly polarized light, the axion electrodynamics has been detected in a 6-SL $MnBi_2Te_4$ in the zero Hall plateau regime[20].

Despite the experimental demonstration of the QAH effect and the axion insulator state, the fabrication of high-quality $MnBi_2Te_4$ devices with expected quantized properties remains a key challenge. In previous experiments, most odd-number-SL $MnBi_2Te_4$ exhibited a small AH effect that is far from quantization[3,21,22], while even-number-SL devices usually exhibited linear normal Hall effect with negative slope in the AFM regime[4,23]. More puzzlingly, recent magnetic and transport measurements[24-27] found that the AH effect disappeared in some odd-number-SL $MnBi_2Te_4$ with uncompensated AFM order, whereas a pronounced AH hysteresis occurred in some even-number-SL devices with fully compensated AFM order. Interestingly, the chirality of the AH hysteresis is opposite to the expected clockwise chirality for Mn-based TIs[28-30]. These counter-intuitive results have aroused widespread controversies regarding the topological nature of $MnBi_2Te_4$ in the AFM state, significantly impeding the explorations of other exotic topological quantum phenomena in topological antiferromagnets[14-16,31]. Several distinct scenarios have been proposed to account for these anomalies, such as the competition between intrinsic and extrinsic mechanisms of AH effect[32], the magnetoelectric effect from the orbital magnetization[33], and the layer-dependent hidden Berry curvature[34]. However, all the ideas assume $MnBi_2Te_4$ crystals with perfect sample qualities and electronic structure. As has been demonstrated by experiments[35-37], even starting with the most optimized crystal, the electronic structure of a fabricated device may change dramatically, which is a critical issue in $Bi_2Te_3$ family TI materials. Theoretical calculations also suggested that the surface defects can result in redistribution of the surface charge from the first layer toward the second layer[38], which will modify the magnetotransport performance of few-layer $MnBi_2Te_4$. Consequently, a promising yet unexplored research direction is to elucidate whether the fabrication process can lead to distinct transport behaviors in $MnBi_2Te_4$, which may offer a novel perspective for resolving the discrepancies in previous experiments.

In this work, we report systematic magnetotransport studies and the evolution of optical contrast ($O_c$) on 223 $MnBi_2Te_4$ devices with varied thickness. All the seven transport devices (from 5 SL to 8 SL) manifest quantized $\rho_{yx} \sim h/e^2$ in the field-polarized Chern insulator state, suggesting the high quality of our $MnBi_2Te_4$ devices. We demonstrate that fabrication process can result in mismatched even-odd layer dependent magnetotransport in few-layer $MnBi_2Te_4$. A comprehensive study of the magnetotransport behaviors in a 6- and 7-SL device shows that both even- and odd-number-SL $MnBi_2Te_4$ can exhibit zero Hall plateau in zero magnetic field. A statistical survey of the $O_c$ in more than 200 $MnBi_2Te_4$ reveals that the effective thickness for magnetotransport could decrease by 1 SL after undergoing the electron-beam-lithography (EBL) method. Our finding not only provides an explanation to the controversies concerning the even-odd discrepancy of magnetotransport in few-layer $MnBi_2Te_4$, but also highlights the critical issues regarding the fabrication and characterization of 2D material devices.

## Results

### Device Fabrication and Basic Calibration of Transport Properties

$MnBi_2Te_4$ few-layer flakes were prepared via mechanical exfoliation on 285 nm $SiO_2$/Si substrates (see methods section). We then determined the thickness of the flakes using optical methods (Fig. 1b), atomic force microscopy (Fig. 1c) and scanning superconducting quantum interference device (SQUID) (see supplementary section A). The calibration of thickness was also examined by additional layer-dependent measurements on flakes exfoliated from crystal #1, including nonlocal transport, scanning microwave impedance microscopy (sMIM), ultra-fast pump-probe reflectivity, and Raman spectroscopy[13,39-41]. By conducting $O_c$ measurement immediately after exfoliation in a glovebox, one can quickly determine the thickness without exposing the sample to the atmosphere. Figure 1d summarizes the one-to-one correspondence between $O_c$ and thickness (SL number), which are highly consistent with the results measured in different crystals by another group[23]. For few-layer $MnBi_2Te_4$, a remarkable feature is that $O_c$ changes its sign from negative to positive when the thickness increases from 6 SL to 7 SL, as guided by the dashed line. After the identification of thickness, the flakes were fabricated into field-effect-transistors by standard EBL method and coated with a layer of Polymethyl Methacrylate (PMMA) for protection (see supplementary section A for details). To study the layer-dependent transport properties, we first measured the temperature ($T$) dependent $\rho_{xx}$ for a 6-SL and 7-SL device (S2 and S6) at $\mu_0 H = 0$, with the Fermi levels ($E_F$s) gated to the charge neutrality points (CNPs). Both the two flakes were derived from crystal #1. At the CNP, the transport is mainly conducted by the topological surface states or edge states. Therefore, both devices exhibit overall insulating behavior and display kink feature at their Néel temperatures ($T_N$s). Compared to $T_N \sim 25$ K for $MnBi_2Te_4$ bulk crystals[5], the $T_N$s for the few-layer devices are suppressed to 20.6 K and 21.6 K, respectively, possibly due to the enhanced fluctuations at lower dimensions.

## Layer-dependent Magnetoelectric Transport Properties for Varied $V_g$

As a layered AFM TI, the most intriguing feature of $MnBi_2Te_4$ is the layer-dependent transport properties. We performed systematic $\mu_0 H$ dependent transport measurements on the two devices at different $V_g$s (see supplementary section B for transport data at various $T$s), as presented in Figs. 2a and 2b. With the application of $V_g$, $E_F$ is continuously tuned from the valence band towards the conduction band, manifested by the slope change of normal Hall effect from positive to negative. For the 6-SL $MnBi_2Te_4$, the most remarkable feature lies in the broad zero Hall plateau in the low-field AFM regime when its $E_F$ is tuned within the band gap. In the panels enclosed by thick magenta boundaries, the zero Hall plateau persists in a wide range of $V_g$ from 36 to 49 V. Meanwhile, $\rho_{xx}$ shows insulating behavior and reaches as high as 4 $h/e^2$. These behaviors are indicative of the axion insulator state in even-number-SL $MnBi_2Te_4$, where the counter-propagating surface Hall currents give rise to a broad zero Hall plateau in $\rho_{yx}$ and a large $\rho_{xx}$ (ref. [5,12,13]). An out-of-plane $\mu_0 H$ drives the system into a Chern insulator at the CNP ($V_g = 42$ V), where $\rho_{yx}$ is quantized in $h/e^2$ and $\rho_{xx}$ drops to zero for $\mu_0 H >$ 6 T. These behaviors are consistent with previous reports on the topological phase transition between axion insulator and Chern insulator in a 6-SL device[5,20].

Figure 2b shows the $\mu_0 H$-dependent $\rho_{yx}$ and $\rho_{xx}$ at various $V_g$s for the 7-SL device, which exhibit unexpected zero Hall plateau phenomenon rather than AH hysteresis in the AFM state. At high field, the 7-SL device show transport behaviors very similar to the 6-SL device with quantized $\rho_{yx}$ and vanished $\rho_{xx}$, as the Chern insulator quantization in the FM state does not depend on thickness. However, in the low-field AFM regime, some unexpected behaviors are observed. As guided by the black dashed lines, throughout the $V_g$ range, $\rho_{yx}$ displays overall linear behaviors and smoothly changes the slope from positive to negative. No discernable hysteresis is observed during the field sweep process. Remarkably, at $V_g = 13$ V, a wide zero Hall plateau appears between $\mu_0 H = \pm 3$ T. Meanwhile, $\rho_{xx}$ reaches the maximum but with a smaller value than that of the 6-SL device. Theoretically, the zero Hall plateau phenomenon is unique to even-number-SL $MnBi_2Te_4$ with fully compensated AFM order, thus should be absent in odd-number-SL $MnBi_2Te_4$. These unexpected results strongly suggest the existence of some unknown mechanism that could modify the magnetotransport of few-layer $MnBi_2Te_4$.

In order to realize the QAH and axion insulator state in few-layer $MnBi_2Te_4$, $E_F$ must be tuned by $V_g$ to lie in the Dirac point gap opened by FM order. To reveal the nature of the zero Hall phenomena in the two devices, we extract the value of $\rho_{xx}$ and the slope of $\rho_{yx}$ at $\mu_0 H = 0$, and plot them as a function of $V_g$. As displayed in Fig. 3a, $\rho_{xx}$ of the 6-SL device first goes up to a large value of 4 $h/e^2$ for $V_g < 25$ V and remains unchanged in a broad $V_g$ window, and then decreases to a small value for $V_g > 50$ V. Meanwhile, $d\rho_{yx}/dH$ exhibits a clear three-stage transition with varying $V_g$. In the first stage with $V_g < 30$ V, $d\rho_{yx}/dH$ progressively decreases with increasing $V_g$, and is attributed to the depletion of hole-type carriers. For $V_g$ from 25 to 30 V, $d\rho_{yx}/dH$ changes sign from positive to negative. As $V_g$ is further increased, a broad zero plateau forms and persists within a $V_g$ window of 13 V. Further application of $V_g$ injects more electron-type carriers and ultimately leads to negative $d\rho_{yx}/dH$. Such behaviors unequivocally suggest that the zero Hall plateau state in the 6-SL $MnBi_2Te_4$ is a genuine quantized Hall state ($C = 0$) with $E_F$ residing in the band gap, which is consistent with our previous report[5].

Despite the superficially similar zero Hall plateau during $\mu_0 H$ sweep in the 7-SL device, it manifests different behavior in response to $V_g$. In contrast to the 6-SL device where $d\rho_{yx}/dH = 0$ exists in a broad $V_g$ window, for the 7-SL device, $d\rho_{yx}/dH = 0$ only appears at a single $V_g$ point corresponding to the sign change of $\rho_{yx}$ slope. Meanwhile, we notice that for the 6-SL device, there is a broad $V_g$ range where the zero Hall plateau and the Chern insulator coexist. However, for the 7-SL device, the zero Hall plateau only occurs in a $V_g$ smaller than the Chern insulator regime (see supplementary section C for colormaps of $\rho_{yx}$ and $\rho_{xx}$). For longitudinal transport, the $V_g$ range for large $\rho_{xx}$ in the 7-SL device is also narrower than the 6-SL device. To better visualize the different manifestations of the zero Hall plateaus, we summarize the variations of $d\rho_{yx}/dH$ with $V_g$ and $\mu_0 H$ for the two devices to two colormaps, as shown in Figs. 3c and 3d. The magenta dashed lines label the regimes for $d\rho_{yx}/dH = 0$. It clearly shows that there is a well-defined zero Hall resistivity plateau regime in the parameter space for the 6-SL device. However, for the 7-SL device, the zero Hall plateau exists in a narrower regime. The quantitative differences of the zero Hall plateaus in the $V_g$ range, as well as that in the $T$ range (see supplementary Fig. S4), indicate different manifestations of the zero Hall plateau associated with the axion insulator state of different energy gaps.

The observation of zero Hall plateau phenomenon in the 7-SL device bears resemblance to a recent observation of the discrepancies between magnetic order and transport properties in few-layer $MnBi_2Te_4$, where the absence of AH effect was observed in a 5-SL device with uncompensated AFM order, meanwhile a pronounced AH effect was found in a 6-SL device with fully compensated AFM order[24,25]. Previous magnetic measurements have demonstrated that the AFM order in $MnBi_2Te_4$ is highly robust and persist to the top surface[42,43]. In contrast, the surface electronic band structures have been found to be fragile and sensitive to the type and concentration of defects[38,44-47]. We notice that most of the magnetization measurements in previous reports[24,25] were performed on $MnBi_2Te_4$ with fresh surface, whereas the transport measurements were conducted exclusively in devices after fabrication. It is highly possible that the even-odd discrepancy of magnetotransport in $MnBi_2Te_4$ arises from the influences of fabrication process. To verify our conjecture, we tracked the $O_c$ values measured before and after fabrication for the two devices, as illustrated in Figs. 3c and 3d. Surprisingly, we find a substantial $O_c$ reduction from +12.5 to -0.2 % for the 7-SL device after fabrication, indicating that the thickness determined by $O_c$ is significantly reduced by 1 SL. In contrast, the $O_c$ value of the 6-SL device is less influenced, only changing slightly from -7.4 to -10.0 %.

**Statistical Survey of Optical Properties and its Effects on Charge Transport**

In order to figure out the reason for the color change and to exclude any artificial factor that may contribute to our observation, such as the transport electrodes, fabrication conditions, and imaging parameters *etc.*, we conducted thorough control experiments on many few-layer flakes and compared $O_c$ changes under different conditions (see supplementary section D for details). To mitigate the potential interferences from extrinsic effects, such as thermal cycling, environmental doping, and aging effect, $O_c$ was obtained immediately after surface treatment in a glovebox[36,37,48,49]. Of the many relevant factors, we notice that the contact with PMMA plays the most crucial role on the reduction of $O_c$, a factor that was rarely noticed in previous studies of 2D materials. We have performed a statistical survey on more than 200 $MnBi_2Te_4$ exfoliated from four crystals grown by different groups, and the main results are summarized in Figs. 4a and 4b. The most striking observation is that most of the studied $MnBi_2Te_4$ flakes exhibit $O_c$ reduction, although to different extents, which is never reported in previous studies of $MnBi_2Te_4$. As presented in Fig. 4b, the blue and magenta dashed lines mark the area of $O_c$ reduction of 0 and 20 %, respectively. The flakes situated close to the blue dashed line display little $O_c$ change after device fabrication, whereas the flakes close to the magenta dashed line experience a pronounced $O_c$ reduction, corresponding to an effective thickness decrease of 1 SL. The subtle increase of $O_c$ in some certain samples is attributed to measurement error (see methods). In the top panel of Fig. 4a, we present the optical images of four typical $MnBi_2Te_4$ flakes, which clearly illustrate the pronounced color change caused in the fabrication process. In Fig. 4c, we further analyze the distribution of the $O_c$ change for the different crystals. The leftward shift of the center of the blue lines clearly indicates that the impacts of the fabrication process on $O_c$ are highly crystal-dependent. For most of the samples exfoliated from Crystal #1, their $O_c$ values are only slightly affected. In contrast, almost all the flakes exfoliated from Crystal #4 exhibit significant reduction in $O_c$, corresponding to a thickness of 1 SL.

## Discussion

Based on the above experimental observations, we discuss the possible explanations for the even-odd discrepancy of magnetotransport in few-layer $MnBi_2Te_4$. It may be suspected that one physical layer is unintentionally removed during the fabrication process, leading to an odd (even)-number-SL $MnBi_2Te_4$ to manifest transport behaviors that are characteristic of an even (odd)-number-SL $MnBi_2Te_4$ with 1 less SL[24,25]. However, such scenario can be safely excluded. We performed atomic force microscopy measurement on the flakes exfoliated from the most sensitive crystal (#4). All these samples exhibit pronounced $O_c$ reduction during the fabrication process (see supplementary section E), however their physical heights determined by atomic force microscopy remain unchanged. In supplementary section F, we also compare the variations in the magneto-optical Kerr effect (MOKE) and the coherent interlayer phonon frequency of two $MnBi_2Te_4$ before and after PMMA contact. It further demonstrates that the fabrication mainly affects the effective thickness rather than the physical thickness. Therefore, a more plausible scenario is that the change of $O_c$ arises from the modification of the magnetic or electronic structures[45,46,50]. In the experimental researches of $MnBi_2Te_4$, it is a widespread phenomenon that $MnBi_2Te_4$ exhibits sample-dependent behaviors, whether between different crystals or different flakes exfoliated from the same crystal[3,47,51]. A prevailing understanding attributes this to the various defects and the non-uniformity within $MnBi_2Te_4$ bulk crystal. It has been highlighted that the surface defects and the perturbations to the surface can result in instability of $MnBi_2Te_4$ (refs. 38,47,50-56). Given the intricate physical and chemical process involved in the fabrication process, we attribute the $O_c$ variation to the fabrication-catalyzed instability of $MnBi_2Te_4$ surface.

It is worth noting that some imaging experiment and theoretical calculations have clearly identified some physical mechanisms that can result in a decrease of effective thickness. For instance, a scanning transmission electron microscopy imaging experiment demonstrated that the synergistic effect of a high concentration of Mn-Bi site mixing and Te vacancy can trigger a surface reconstruction process from one SL of $MnBi_2Te_4$ to a quintuple layer of Mn-$Bi_2Te_3$ and an amorphous double layer of $Mn_xBi_yTe$ (ref. 50). As a result, the effective thickness for the $MnBi_2Te_4$ structure is reduced by 1-SL. Theoretical calculations also reveal that a surface charge redistribution process can relocate the surface state from the first SL to the second SL, resulting in the decrease of effective thickness for magnetotransport[51]. Recently, a theoretical work demonstrates that a small expansion of the interlayer van der Waals gap can result in a noteworthy reduction in the surface gap[56]. Specifically, for a (7+1) SL $MnBi_2Te_4$, it triggers a topological phase transition with Chern number change by one. An odd (even)-number-SL $MnBi_2Te_4$ will naturally manifest magnetotransport properties akin to its even (odd)-number-SL counterpart with 1 less SL. Based on the sample-dependent defect type and concentration, as well as the susceptibility of $MnBi_2Te_4$ surface to perturbations[47,48,50,51,56], we hypothesize that the sample dependent behaviors observed during the fabrication arise from the PMMA-catalyzed surface instability. Notably, prior researches on graphene, $MoS_2$, and $WSe_2$ indeed suggested that the PMMA residuals on the surface influence the intrinsic properties of the 2D materials[57-59]. It can not only increase the observed thickness in the atomic force microscopy measurement through absorption, but also act as charge source, prompting the surface charge redistribution. Our topography measurement has indeed shown island-like PMMA residuals on the $MnBi_2Te_4$ surface (see supplementary Fig. S8). In addition, various adsorbates trapped between layers during the fabrication can also expand the van der Waals gap[60]. Therefore, it is likely that the combined influences of non-uniformity, defects, and PMMA contribute to the sample dependent behaviors in response to fabrication. Further studies are needed to fully understand the underlying mechanisms. In Figs. 4d and 4e, we display the process of effective thickness reduction with the magenta frame indicating the effective thickness for transport. The reduced gap elucidates the narrower $V_g$ and $T$ range of the zero Hall plateau for the 7-SL sample (S6).

While the precise mechanism through which PMMA influences the quality of $MnBi_2Te_4$ samples remains incompletely understood, a potential solution to circumvent such fabrication issue involves isolating PMMA from the surface during the fabrication. Building upon recent advancements in low-damage lithography in the QAH system[35,37], we suggest that depositing a thin layer of $AlO_x$ on the surface of $MnBi_2Te_4$ prior to fabrication may alleviate the damage of PMMA. In supplementary section G, we present our preliminary results obtained in crystal #5, which demonstrates the efficacy of the modified method in addressing the current issue.

In addition to the zero Hall plateau in the 7-SL $MnBi_2Te_4$ device, the fabrication-induced mismatched layer dependent magnetotransport behaviors are also evident in $MnBi_2Te_4$ flakes with other thicknesses, as displayed in Figs. 4f and 4g. Among the seven samples, devices S1 and S5 were derived from crystals #3 and #2, respectively. All the other devices were derived from crystal #1. Notably, those PMMA-insensitive $MnBi_2Te_4$ with less-affected $O_c$ (blue stars in Fig. 4b) exhibit the anticipated behaviors for both even- and odd-number-SL $MnBi_2Te_4$. In contrast, samples with pronounced $O_c$ change (red stars in Fig. 4b) exhibit transport behaviors inconsistent with their nominal thickness. Specifically, as shown in Fig. 4g, odd-number-SL devices display vanished AH hysteresis in the AFM regime, while even-number-SL devices display hysteresis behaviors with counterclockwise chirality, as indicated by the black arrows. The AH effect with reversed chirality may arise from the electric field due to gate or substrate, or the competition between various intrinsic and extrinsic mechanisms[23,32,33,48,61]. In addition to the Hall effect, since the transport of odd- and even-number-SL $MnBi_2Te_4$ are conducted by chiral and helical edge states[13,39], the fabrication-induced mismatched even-odd dependent magnetotransport should also be manifested by the nonlocal transport measurements, which are observed in our experiment (see supplementary section H for details).

We have conducted a comprehensive investigation of the transport properties in a large number of few-layer $MnBi_2Te_4$ flakes. By tracking the quantized Hall plateau with respect to $\mu_0 H$ and $V_g$, and comparing the optical properties before and after the fabrication process, our study elucidates the relationship between transport behaviors and device fabrication process. Our research has uncovered a condition in which the effective thickness for charge transport in $MnBi_2Te_4$ becomes decoupled from its pristine physical thickness, which is never reported in previous studies. Although the exact microscopic mechanism underlying the change of $O_c$ remains to be determined, and we cannot exclude that those devices exhibiting unchanged $O_c$ are not affected by fabrication because the AH effect (0.1 $h/e^2$) in odd-number-SL $MnBi_2Te_4$ is not quantized, our experiments still provide highly valuable insights for the fabrication of high-quality $MnBi_2Te_4$ toward realizing quantized phenomena. Our finding not only explains the controversies concerning the mismatched even-odd layer dependent magnetotransport in $MnBi_2Te_4$, but also highlights the critical issues regarding the fabrication and characterization of devices based on 2D materials.

## Methods

**Crystal growth** High-quality $MnBi_2Te_4$ single crystals were synthesized independently by different methods. For crystal #1, it was grown by directly mixing $Bi_2Te_3$ and MnTe with the ratio of 1:1 in a vacuum-sealed silica ampoule. After heated to 973 K, the mixture was slowly cooled down to 864 K, followed by a long period of annealing process. The phase and crystal structure were examined by X-ray diffraction on a PANalytical Empyrean diffractometer with Cu Kα radiation. For crystal #2, it was grown by conventional flux method. Mn powders, Bi and Te were weighed with the ratio Mn:Bi:Te = 1:8:13 (MnTe:$Bi_2Te_3$ = 1:4) in an argon-filled glovebox. The mixtures were loaded into a corundum crucible which was sealed into a quartz tube. Then the tube was then put into a furnace and heated up to 1000 °C for 20 hours. After a quick cooling to 605 °C with the rate of 5 °C/h, the mixtures were then slowly cooled down to 590 °C with the rate of 0.5 °C/h and kept for 2 days. Finally, the crystals were obtained after centrifuging. For crystal #3, it was grown by the conventional high-temperature solution method. The Mn, Bi and Te blocks were weighed with a ratio of Mn:Bi:Te = 1:11.3:18, and placed in an alumina crucible, which were then sealed in a quartz tube in argon environment. The assembly was first heated up in a box furnace to 950 °C and held for 10 hours, and then cooled down to 700 °C within 10 hours and further cooled down to 575 °C in about 100 hours. After the heating procedure, the quartz tube was then taken out quickly and decanted into the centrifuge to remove the flux from the crystals. For crystal #4, it was grown by flux method using $MnCl_2$ as the flux. The raw materials of $Bi_2Te_3$ powder, Mn lump, Te lump and $MnCl_2$ powder were mixed with a molar ratio of 1:1:1:0.3 and then placed in a dry alumina crucible, which was sealed in a fused silica ampoule under vacuum. The ampoule was then placed in a furnace and heated up to 850 ℃ for over 20 hours, kept there for 24 hours, cooled down to 595 ℃ in over 5 hours, kept there for 150 hours, and finally cooled to room temperature in 5 hours. After the steps above, the yielded ingot was cleaved into millimeter-sized crystals with metallic luster. For crystal #5, it was grown by directly mixing $Bi_2Te_3$, MnTe and Te with the ratio of 1:1:0.2 in a vacuum-sealed silica ampoule. The ampoule was slowly heated to 900°C at a rate of 3°C/min and maintained at this temperature for 1 hour. Subsequently, the sample was cooled at a rate of 3°C/min to 700°C, held at this temperature for 1 hour. The temperature was then gradually decreased to 585°C at a rate of 0.5°C/min and maintained for annealing for 12 days. After the annealing process, the quartz ampoule was quenched in water to avoid phase impurities. Millimeter-sized $MnBi_2Te_4$ crystals were obtained after crushing the ingot.

**Device fabrication** $MnBi_2Te_4$ flakes were exfoliated onto 285 nm-thick $SiO_2$/Si substrates by using the Scotch tape method in an argon-filled glove box with $O_2$ and $H_2O$ levels lower than 0.1 ppm. Before exfoliation, all $SiO_2$/Si substrates were pre-cleaned by air plasma for 5 minutes at ~ 125 Pa pressure. To minimize the experimental errors due to the subtle difference in measurement conditions, such as the position of the flakes in the light fields, the uniformity of illumination, the size and shape of the sample, and the presence of electrode, the $O_c$ shown in the main text were calculated by averaging the $O_c$ of different parts across the sample. For the transport devices, thick flakes around the target sample were first scratched off by using a sharp needle in the glove box. A layer of 270 nm PMMA was spin-coated before EBL and heated at 60 °C for 5 minutes. After the EBL, 23 to 53 nm thick Cr/Au electrodes (3/20 to 3/50 nm) were deposited by a thermal evaporator connected with an argon-filled glove box. Before the fabrication and sample transfer process, the devices were always spin-coated with a PMMA layer to avoid contact with air. All the seven devices (S1-S7) shown in the text were fabricated through the same process.

**Transport measurement** Four probe transport measurements were carried out in a cryostat with the lowest temperature 1.6 K and out-of-plane magnetic field up to 9 T. The longitudinal and Hall signals were acquired simultaneously via lock-in amplifiers with an AC current (200 nA, 13 Hz) generated by a Keithley 6221 current source meter. To correct for the geometrical misalignment, the longitudinal and Hall signals were symmetrized and antisymmetrized with magnetic field respectively. The back-gate voltages were applied by a Keithley 2400 source meter.

**Scanning SUIID measurement** Scanning SQUID measurements were carried in a different cryostat from the transport measurements. Scanning 2-junction SQUID susceptometers with two balanced pickup loops of 2 μm diameter in a gradiometric configuration were utilized as the SQUID sensors. Each of them was surrounded by a one-turn field coils of 10 μm diameter. The DC flux was measured through the pickup loop using a voltage meter (Zurich Instrument HF2LI) as a function of position and reflects the intrinsic magnetization of the sample.

**Polar MOKE measurement** Polar MOKE measurements were carried using a 633 nm HeNe laser. After transmitting through a linear polarizer, the light was focused to a 2μm spot on the sample by a reflective objective at normal incidence to avoid the large backgrounds that occur when a typical lens is used. The sample was mounted on a cold stage at 3 K within the vacuum chamber of an optical superconducting magnet system. The reflected beam is modulated at ~ 50 kHz by a PEM, split by a Wollaston prism, and detected using a balanced photodiode. The resulting 50 and 100 kHz modulations detected by lock-in amplifiers then correspond to the ellipticity and rotation angle of the beam respectively. We additionally modulate the intensity of the beam with a frequency of 2317 Hz chopper to measure the DC signal for normalization using a third lock-in.

**Data Availability:** All data supporting the finding in the study are presented within the main text and the supplementary information. All data are available upon reasonable request from the corresponding author.

**Acknowledgements:** Chang Liu (RUC) was supported by fundings from National Natural Science Foundation of China (Grant No. 12274453) and Open Research Fund Program of the State Key Laboratory of Low-Dimensional Quantum Physics (Grant No. KF202204). Jinsong Zhang was supported by funding from National Natural Science Foundation of China (Grants No. 12274252 and No. 12350404). Yayu Wang was supported the Basic Science Center Project of Natural Science Foundation of China (Grant No. 52388201), the New

Cornerstone Science Foundation through the New Cornerstone Investigator Program and the XPLORER PRIZE. Yayu Wang, Jinsong Zhang, and Chang Liu (RUC) acknowledge the financial support from Innovation program for Quantum Science and Technology (Grant No. 2021ZD0302502). Yang Wu was supported by funding from National Natural Science Foundation of Chinse (Grants No. 51991340 and No. 51991343). Shuang Jia was supported by fundings from the National Natural Science Foundation of China (Grants No. 12225401 and No. 12141002), the National Key Research and Development Program of China Grant No. 2021YFA1401902. Chang Liu (SUSTC) was supported by funding from the National Natural Science Foundation of China (Grant No. 12074161). Tianlong Xia was supported by fundings from the National Natural Science Foundation of China (Grant No. 12074425), the National Key R&D Program of China (Grant No. 2019YFA0308602), and the Fundamental Research Funds for the Central Universities, and the Research Funds of Renmin University of China (No. 23XNKJ22). Yihua Wang acknowledge support by National Key R&D Program of China (Grant No. 2021YFA1400100), National Natural Science Foundation of China (Grant No. 12150003) and Shanghai Municipal Science and Technology Major Project (Grant No. 2019SHZDZX01). Luyi Yang was supported by fundings from the National Natural Science Foundation of China (Grants No. 12074212 and No. 12361141826).

**Author contributions:** C. L. (RUC), Y. Y. W. and J. S. Zhang supervised the research. C. L. (RUC), Y. X. L., Y. C. W., Z. C. L., L. C. X., Y. F., Y. Q. W, B. H. F, and S. Y. fabricated the devices and performed the transport measurements. Y. C. W., H. L., Y. W., H. W., T. L. X., R. E. L., C. L. (SUST), L. F. L and S. J. grew the $MnBi_2Te_4$ crystals. Y. C. W., Z. C. L., and Z. T. G performed the atomic force microscopy measurements. J. J. Z., Y. F. and Y. H. W. performed the scanning SQUID measurements. L. Y. L. and L. Y. Y. carried the MOKE measurements. C. L. (RUC), Y. X. L. and Y. Y. W. prepared the manuscript with comments from all authors.

**Competing interests:** The authors declare no competing interests.

## Figure Captions

**Fig. 1 | Crystal structure and basic calibration of few-layer $MnBi_2Te_4$. a**, Crystal structure

of an even-number-SL $MnBi_2Te_4$. **b**, Optical image of few-layer $MnBi_2Te_4$ flakes exfoliated on $SiO_2$/Si substrate. The numbers in the figure represent the $O_c$ and the corresponding layer nubmers. Here $O_c$ is defined as $(I_{flake} - I_{substrate})/I_{substrate}$, where $I_{flake}$ and $I_{substrate}$ are the intensity of $MnBi_2Te_4$ flake and substrate. **c,** Atomic force microscope morphology of the area marked by the dashed box in **b** and height profile of the $MnBi_2Te_4$ along the red line. **d,** Variation of $O_c$ as a function of thickness. The numbers in parentheses represent the quantity of measured samples. The error bar is defined by the standard deviation of multiple measurements of the data. **e,** $T$ dependence of $\rho_{xx}$ for the 6- and 7-SL device measured at $\mu_0 H = 0$ when $E_F$s are gated to the CNPs. The red and blue arrows mark the $T_N$s for AFM transition.

**Fig. 2 | Distinct evolution of the zero Hall plateau for 6- and 7-SL $MnBi_2Te_4$ devices. a-b,** $V_g$ dependent $\rho_{xx}$ and $\rho_{yx}$ for the 6-SL (**a**) and 7-SL (**b**) device measured at $T = 1.5$ K. The black dashed lines denote the slope of $\rho_{yx}$ in low-field AFM regime. For the 6-SL device, the zero Hall plateau exists in a wide $V_g$ range from 36 to 49 V, while for the 7-SL device it only occurs in specific $V_g$ around 13 V. The panels enclosed by thick magenta boundaries indicate the $V_g$ regime where zero Hall phenomena exist.

**Fig. 3 | $V_g$ dependent transport properties and optical images for 6- and 7-SL $MnBi_2Te_4$ devices. a-b,** $\rho_{xx}$ (blue) and $d\rho_{yx}/dH$ (red) at $\mu_0 H = 0$ as a function of $V_g$ for the 6-SL (**a**) and 7-SL (**b**) device. For the 6-SL device, $d\rho_{yx}/dH$ changes sign from positive to negative at $V_g$ = 30 V, followed by a broad zero Hall plateau with a range of 13 V. Whereas for the 7-SL device, $d\rho_{yx}/dH$ monotonously decreases and crosses zero at $V_g$ = 13 V. There is no plateau formation near $d\rho_{yx}/dH = 0$. **c-d,** Colormaps of $d\rho_{yx}/dH$ as functions of $\mu_0 H$ and $V_g$. The magenta dashed lines represent the zero Hall plateau regimes for the two devices. **e-f,** Optical images for the two devices acquired immediately after exfoliation and after fabrication. For the 6-SL device, $O_c$ is less affected during the fabrication process, but for the 7-SL device $O_c$ is significantly reduced from 12.5 to -0.2 %. The different colors of the electrodes in the two devices are due to the different thickness of Au, which does not affect conclusion on the effect of fabrication on $O_c$.

**Fig. 4 | Statistical analysis of $O_c$ for more than two hundred flakes and distinct thickness dependent transport properties. a,** Optical images of four representative samples taken in a glove box right after exfoliation (top panel) and after the removal of PMMA (bottom panel). **b,** Summary of the $O_c$ values of 223 $MnBi_2Te_4$ flakes after exfoliation and after the removal of PMMA. The blue and magenta dashed lines mark the $O_c$ reduction by 0 and 20 %. Different colored dots represent the data acquired from different crystals. **c,** Distribution of $O_c$ change in the four different crystals. For the most PMMA-sensitive crystal (#4), fabrication can give rise to $O_c$ change corresponding to a thickness of 1 SL. **d-e,** Illustrations of the influence of PMMA on the surface electronic structure for a 7-SL $MnBi_2Te_4$. **f-g,** Thickness dependent $\rho_{yx}$ behaviors for $MnBi_2Te_4$ without (blue) and with (red) severe $O_c$ change.

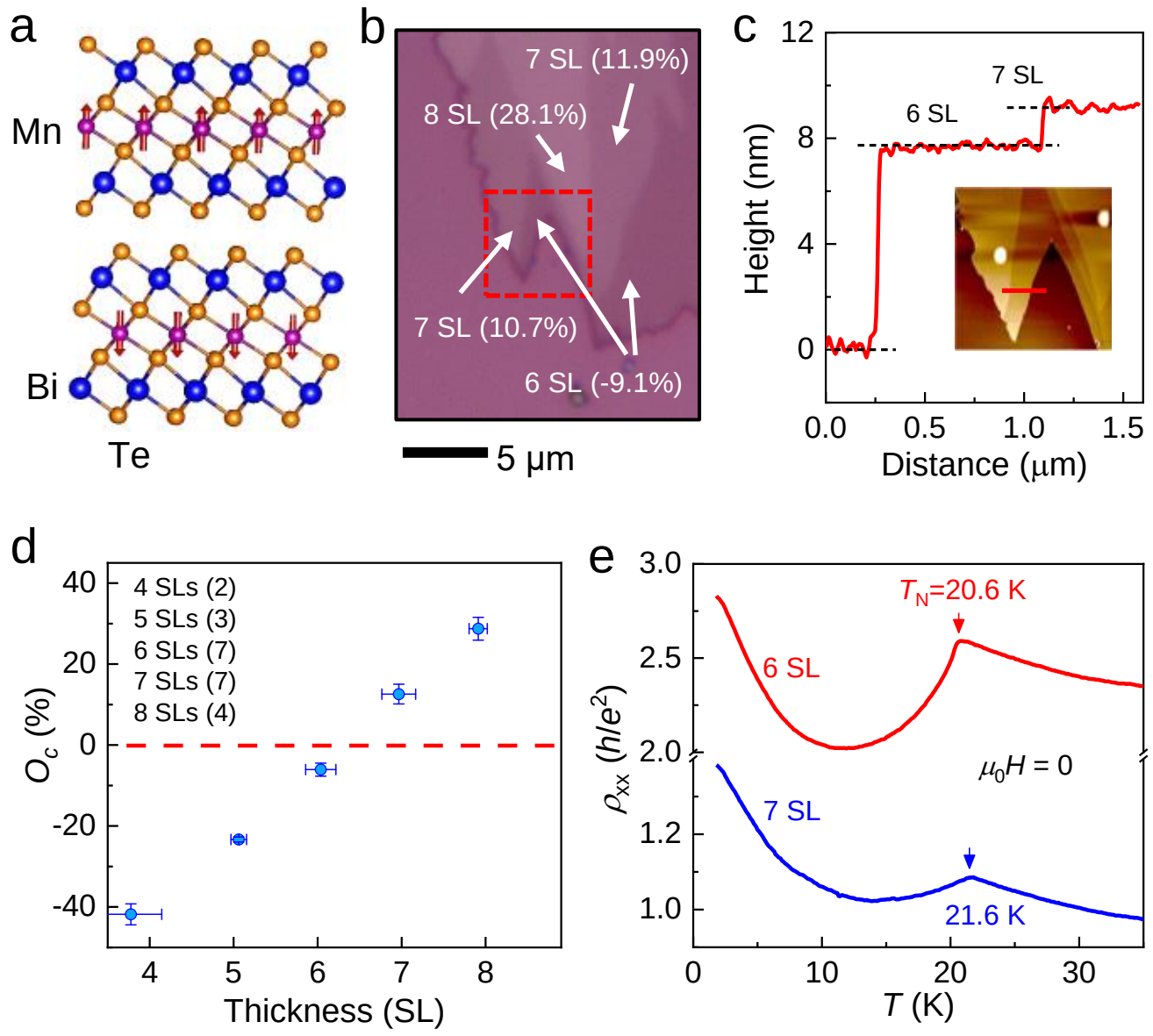

Figure 1

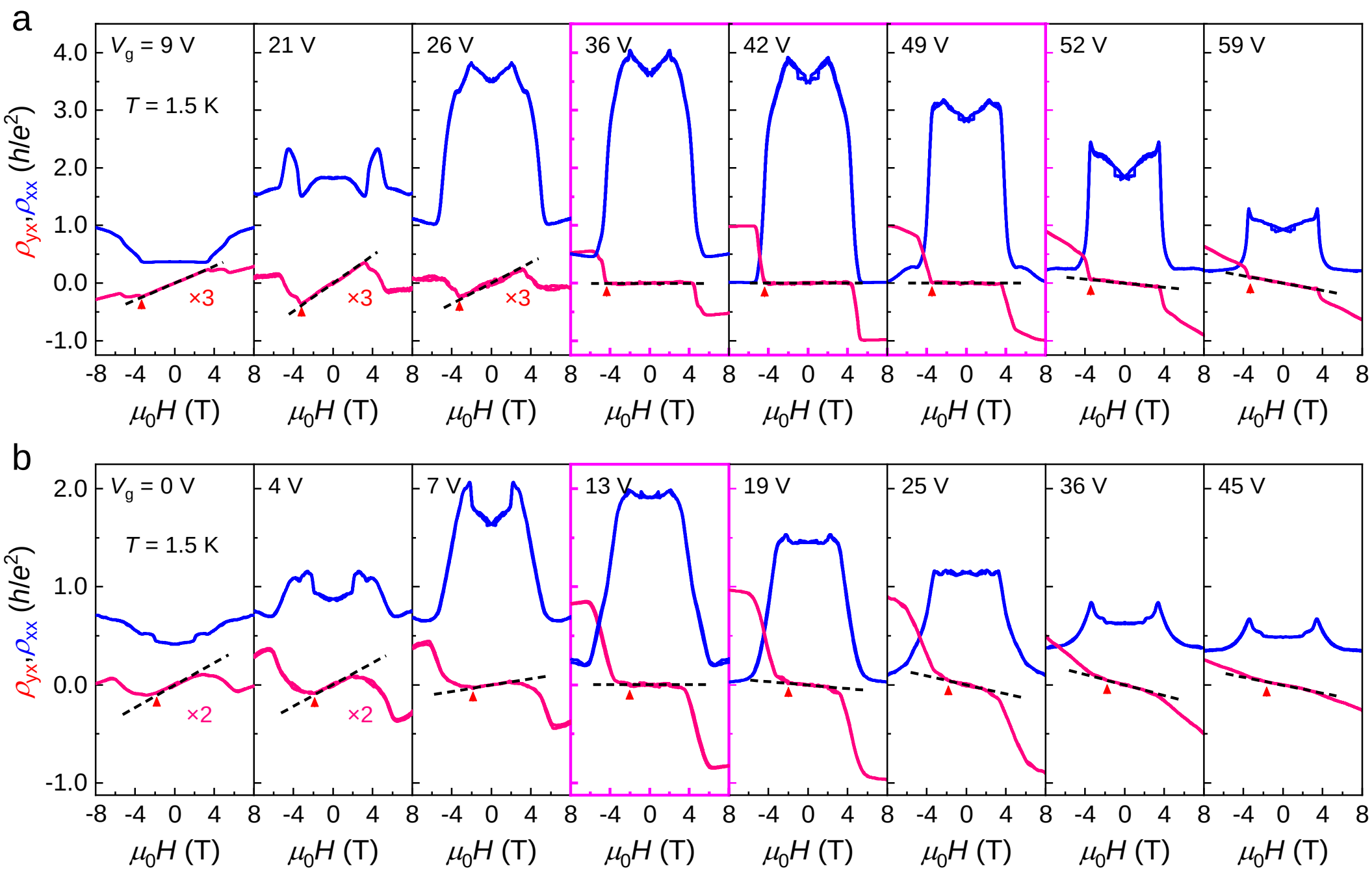

Figure 2

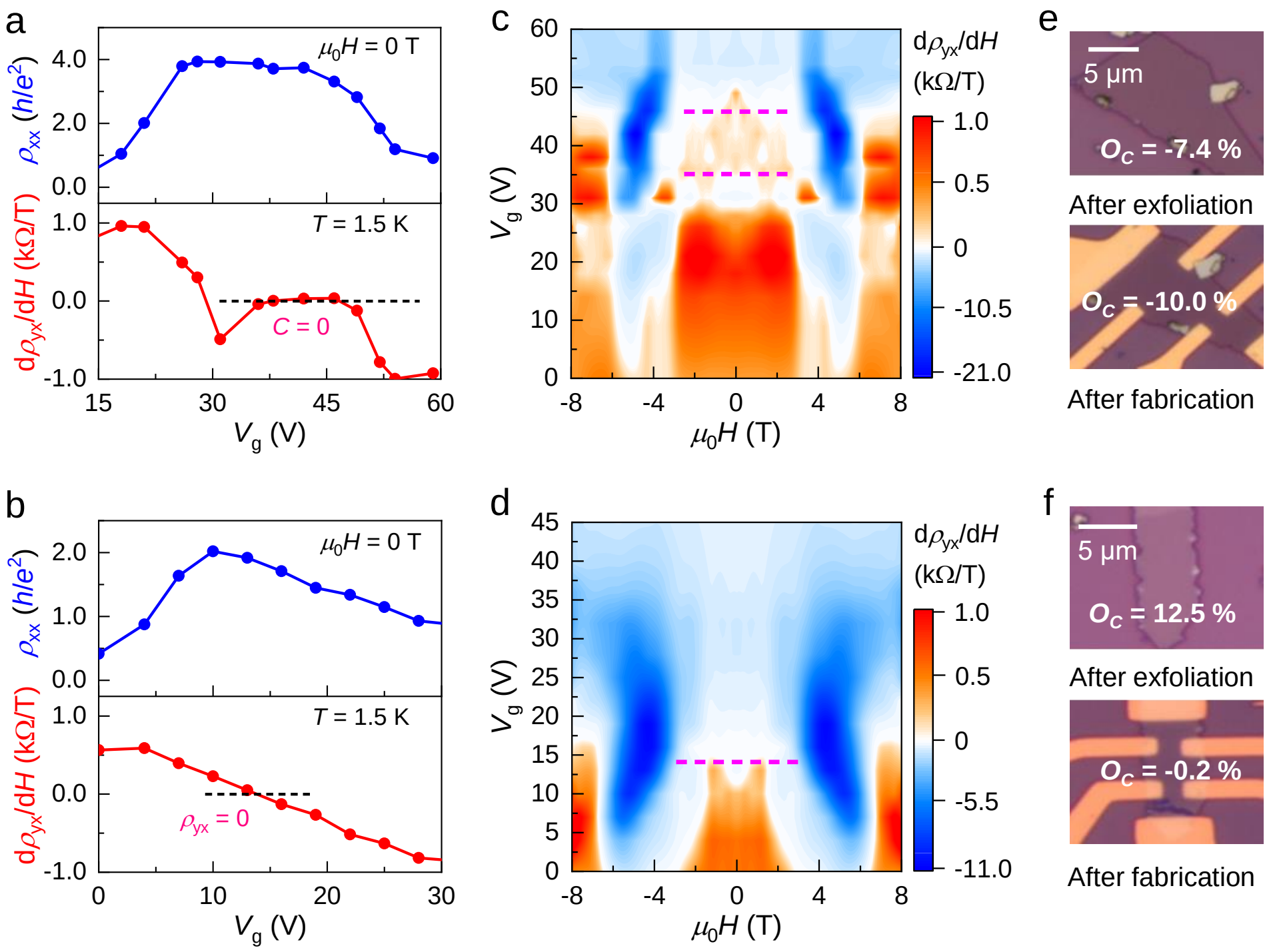

Figure 3

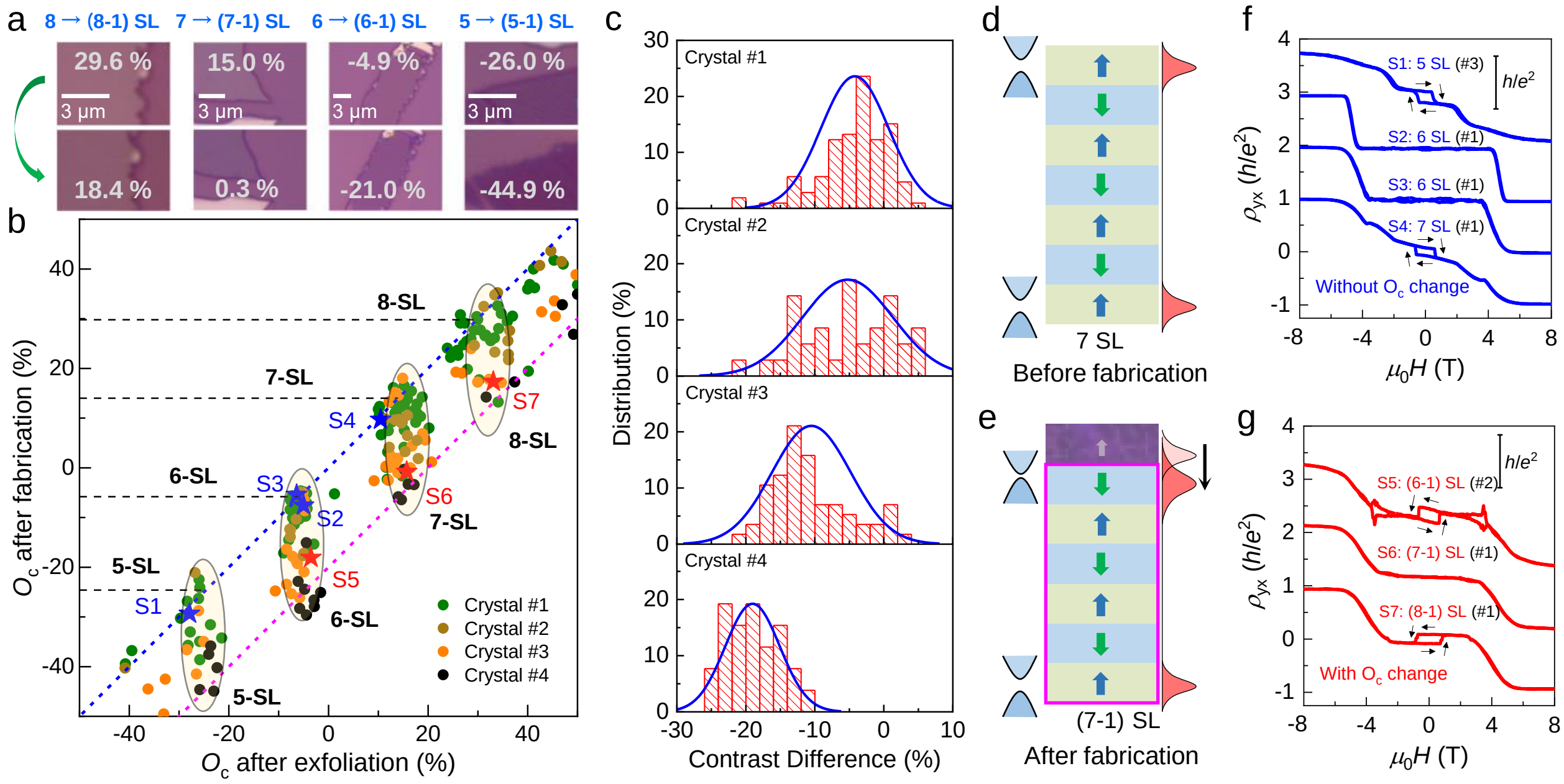

a
8 → (8-1) SL
7 → (7-1) SL
6 → (6-1) SL
5 → (5-1) SL
29.6 %
15.0 %
-4.9 %
-26.0 %
18.4 %
0.3 %
-21.0 %
-44.9 %
3 μm
b
Oc after fabrication (%)
Oc after exfoliation (%)
8-SL
7-SL
6-SL
5-SL
S1
S2
S3
S4
S5
S6
S7
Crystal #1
Crystal #2
Crystal #3
Crystal #4
c
Crystal #1
Crystal #2
Crystal #3
Crystal #4
Distribution (%)
Contrast Difference (%)
d
7 SL
Before fabrication
e
(7-1) SL
After fabrication
f
S1: 5 SL (#3)
S2: 6 SL (#1)
S3: 6 SL (#1)
S4: 7 SL (#1)
Without Oc change
ρyx (h/e2)
μ0H (T)
g
S5: (6-1) SL (#2)
S6: (7-1) SL (#1)
S7: (8-1) SL (#1)
With Oc change

Figure 4